# Influence of structural disorder on low-temperature behavior of penetration depth in electron-doped high-$T_C$ thin films

A. J. C. Lanfredi[a], S. Sergeenkov[b], and F. M. Araujo-Moreira[a,*]

[a] *Grupo de Materiais e Dispositivos, Centro Multidisciplinar para o Desenvolvimento de Materiais Cerâmicos - CMDMC, Departamento de Física, Universidade Federal de São Carlos - UFSCar, Caixa Postal 676, CEP 13565-905, São Carlos - São Paulo, Brasil*

[b] *Bogoliubov Laboratory of Theoretical Physics, Joint Institute for Nuclear Research, 141980 Dubna, Moscow Region, Russian Federation*

**Abstract**

To probe the influence of structural disorder on low-temperature behavior of magnetic penetration depth, $\lambda(T)$, in electron-doped high-$T_C$ superconductors, a comparative study of high-quality $Pr_{1.85}Ce_{0.15}CuO_4$ (PCCO) and $Sm_{1.85}Ce_{0.15}CuO_4$ (SCCO) thin films is presented. The $\lambda(T)$ profiles are extracted from conductance-voltage data using a highly-sensitive home-made mutual-inductance technique. The obtained results confirm a *d*-wave pairing mechanism in both samples (with nodal gap parameter $\Delta_o/k_BT_C$=2.0 and 2.1 for PCCO and SCCO films, respectively), substantially modified by impurity scattering (which is more noticeable in less homogeneous SCCO films) at the lowest temperatures. More precisely, $\Delta\lambda(T)=\lambda(T)-\lambda(0)$ is found to follow the Goldenfeld-Hirschfeld interpolation formulae $\Delta\lambda(T)/\lambda(0)=AT^2/(T+T_0)$ with $T_0=\ln(2)k_B\Gamma^{1/2}\Delta_0^{\ 1/2}$ being the crossover temperature which demarcates pure and impure scattering processes ($T_0/T_C$ =0.13 and 0.26 for PCCO and SCCO films, respectively). The value of the extracted impurity scattering rate $\Gamma$ correlates with the quality of our samples and is found to be much higher in less homogeneous films with lower $T_C$.

* Corresponding author; e-mail address: faraujo@df.ufscar.br

## 1. Introduction

The accurate experimental determination of the temperature behavior of the magnetic penetration depth, λ(T), has been of great interest to the scientific community since the very discovery of high-$T_C$ superconductors. Since the effective value of λ(T) is extremely sensitive to local inhomogeneities, a thorough analysis of its low-temperature profile gives probably one of the most reliable methods to determine the quality of a superconducting material (especially in the form of thin films), which is of utter importance for applications [1,2].

On the other hand, the magnetic penetration depth is strongly sensitive to the variations of the macroscopic superconducting order parameter and therefore its study can give important information about both the symmetry of the superconducting state and the pairing mechanisms.

It is well established that most of the conventional low-$T_C$ superconductors have *s*-wave pairing symmetry. As for high-$T_C$ cuprates, the study of pairing symmetry in these materials has been (and still remains) one of the most polemical and active fields of research over the last few years [2] and the experimental determination of the temperature dependence of λ has been one of the most common methods in these studies. In particular, a power-like dependence $T^n$ of the penetration depth at low temperatures clearly points at a nodal structure of the superconducting gap (as expected for strongly correlated materials) where the exponent n depends on the type of the node in the **k**-space. Most phase-sensitive measurements [3,4] have revealed that hole-doped high-$T_C$ cuprates with nearly optimal doping have predominantly $d_{x^2-y^2}$ pairing symmetry. Regarding the case of the hole-doped cuprate $YBa_2Cu_3O_{7-\delta}$, some groups have reported experimental evidences for a pairing symmetry transition from pure $d_{x^2-y^2}$ (for under-doped compositions) to a mixed-type $d+id_{xy}$ (for over-doped compositions) [5]. At the same time, for electron-doped cuprates, which have composition of the form $Ln_{2-x}Ce_xCuO_4$ (where Ln corresponds to Pr, Nd, or Sm), the pairing mechanisms are not yet fully understood [6-10]. For example, using the point contact spectroscopy technique, Biswas *et al*. [7] have found strong evidences in favor of *d*-wave pairing symmetry in under-doped (x ≈ 0.13) $Pr_{2-x}Ce_xCuO_4$ (PCCO). Further studies revealed [8] that the low temperature superfluid density of Ce-based magnetic superconductors varies quadratically with temperature in the whole range of doping, in agreement with the theoretical prediction for a *d*-wave superconductor with impurity scattering. In addition, remeasured [9] magnetic-field dependence of the low-temperature specific heat of optimally-doped (x=0.15) PCCO give further evidence in favor of *d*-wave-like pairing symmetry in this material at all temperatures below 4.5 K. And finally, the recent penetration depth measurements on $Sm_{1.85}Ce_{0.15}CuO_4$ (SCCO) single crystals [10] have indicated that this magnetic superconductor exhibits a rather strong enhancement of diamagnetic screening below 4 K most probably driven by the Neel transition of Sm sublattice due to spin-freezing of Cu spins.

In this paper we study the influence of local inhomogeneities on low-temperature dependence of the magnetic penetration depth $\lambda(T)$ in high-quality optimally-doped $Pr_{1.85}Ce_{0.15}CuO_4$ (PCCO) and $Sm_{1.85}Ce_{0.15}CuO_4$ (SCCO) thin films grown by the pulsed laser deposition (PLD) technique. The $\lambda(T)$ profiles have been extracted from conductance-voltage data by using a highly-sensitive home-made mutual-inductance bridge.

## 2. Samples Characterization and Experimental Procedure

The structural quality of our samples was verified through X-ray diffraction (XRD) and scanning electron microscopy (SEM) together with energy dispersive spectroscopy (EDS) technique. Both XRD spectra and SEM data reveal that PCCO film is of higher structural quality than SCCO film (see Figures 1 and 2).

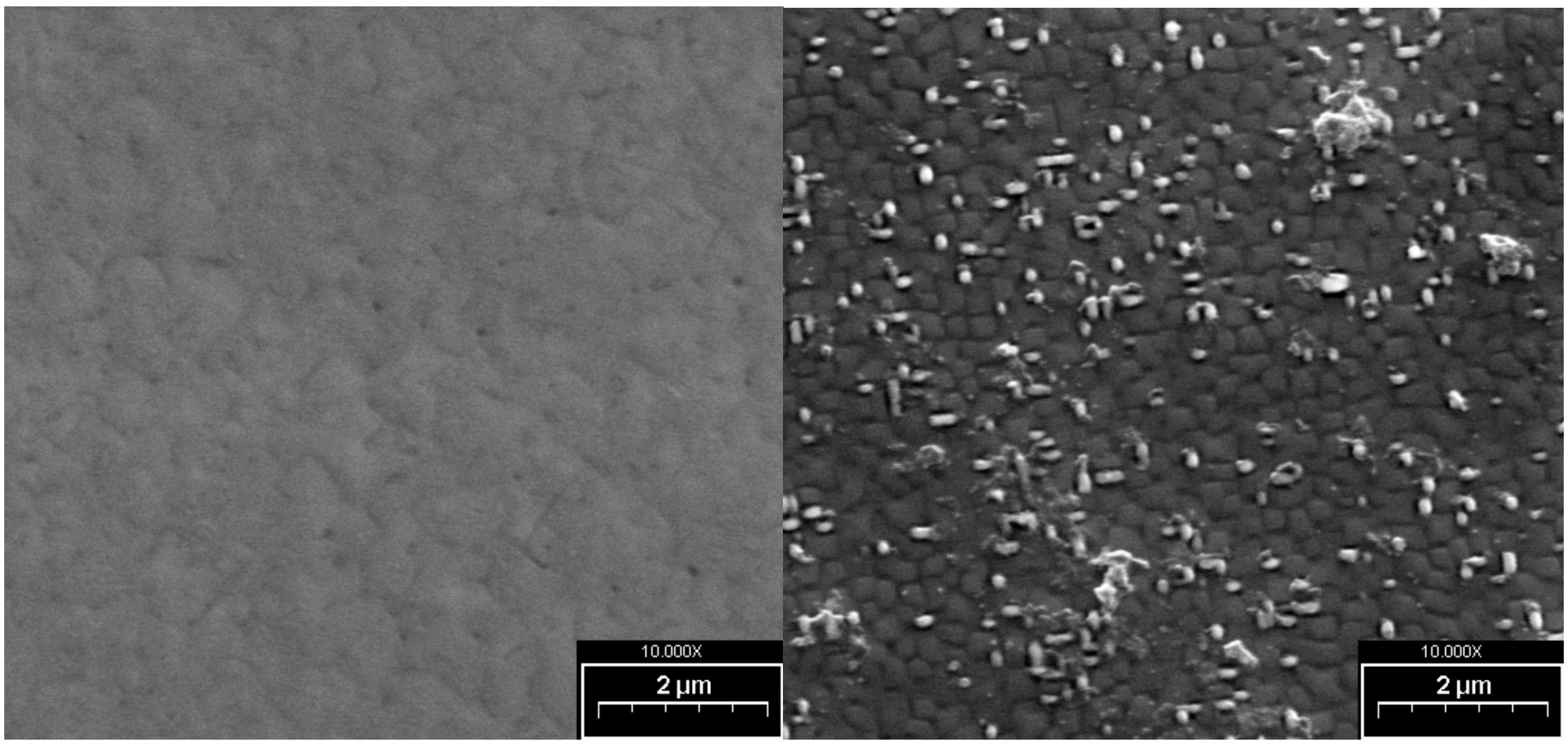

**Figure 1**. SEM scan photography of PCCO (left) and SCCO (right) samples (magnification 10000 times).

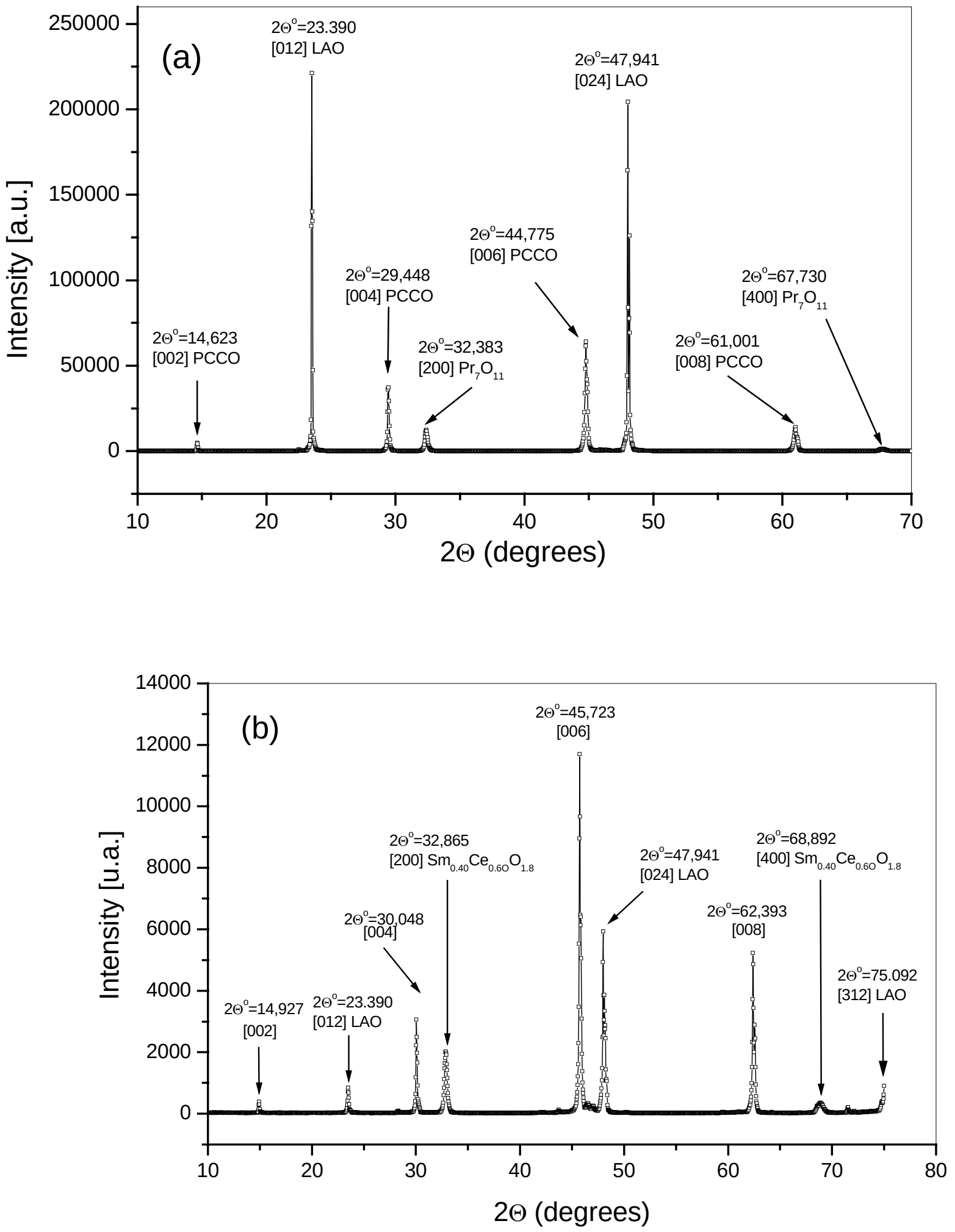

**Figure 2**. X-ray diffraction spectrum of PCCO (a) and SCCO (b) films.

The experimental bridge used in this work is based on the mutual-inductance method. To measure samples in the shape of thin films, the so-called *screening method* has been developed [11]. It involves the use of primary and secondary coils, with diameters smaller than the dimension of the sample. When these coils are located near the surface of the film, the response (i.e., the complex voltage output $V_{AC}$) does not depend on the radius of the film or its properties near the edges. In the reflection technique [12], an excitation (primary) coil coaxially surrounds a pair of counter-wound (secondary) pick-up coils. If we take the current in the primary coil as a reference, $V_{AC}$ can be expressed via two orthogonal components, i.e., $V_{AC} = V_L + iV_R$. The first one is the inductive component, $V_L$ (which is in phase with the time-derivative of the reference current) and the second one is the quadrature resistive component, $V_R$ (which is in phase with the reference current). It can be easily demonstrated that $V_L$ and $V_R$ are directly related to the average magnetic moment and the energy losses of the sample, respectively [13]. When there is no sample in the system, the net output from the secondary coils is close to zero because the pick-up coils are identical in shape but are wound in opposite directions. The sample is positioned as close as possible to the set of coils, to maximize the induced signal in the pick-up coils.

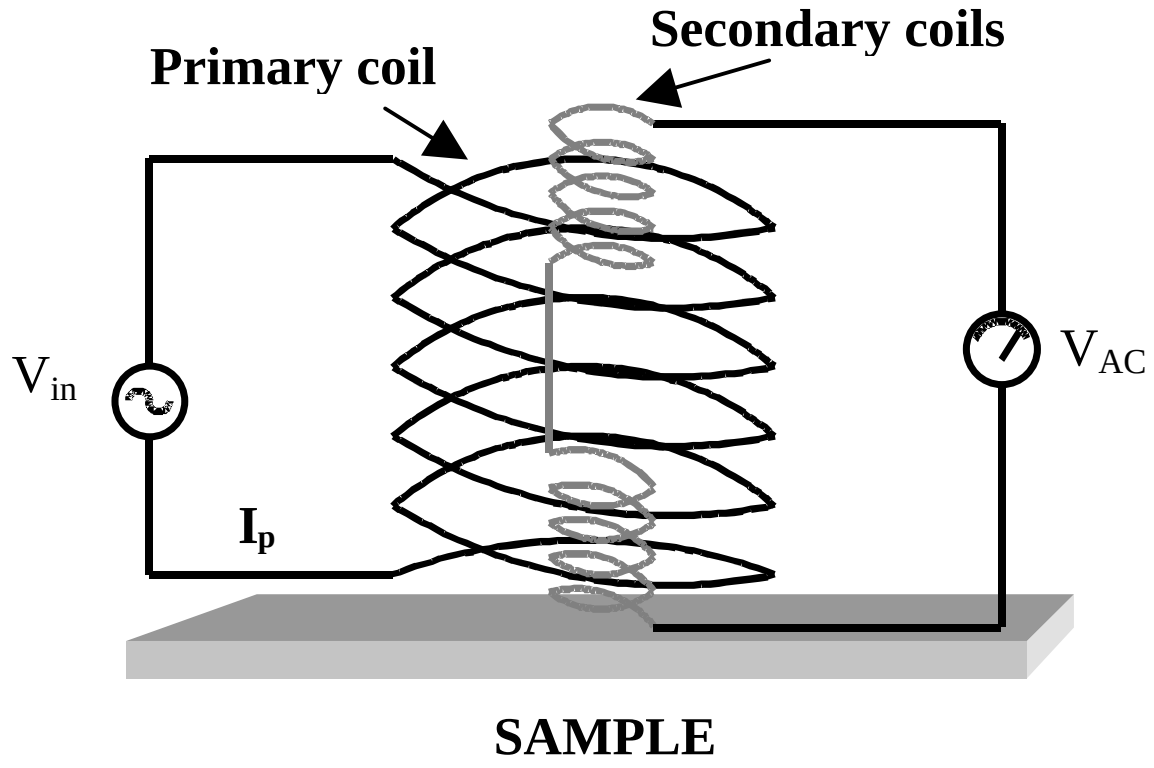

**Figure 3 -** Sketch of the experimental bridge based on the mutual-inductance screening method.

An alternate current sufficient to create a magnetic field of amplitude $h_{AC}$ and frequency f is applied to the primary coil by an alternating voltage source, $V_{in}$. The output voltage of the secondary coils $V_{AC}$ is measured through the usual lock-in technique [14]. Figure 3 shows the sketch of the experimental bridge used in our study based on the mutual-inductance screening method.

## 3. Results and Discussion

To extract the profile of the penetration depth within the discussed here method, one should resolve the following equation relating the measured output voltage $V_{AC}$ to the $\lambda(T)$ sensitive sample features [12]:

$$V_{AC} = V' + iV'' = i\omega I_P \int_0^{\infty} \frac{M(x)}{1 + \frac{2}{\mu_0 (h_P + h_S)} \cdot \frac{1}{i\varpi G} x} \cdot dx \qquad (1)$$

where $I_P$ and $\omega=2\pi f$ are respectively the amplitude and the frequency of the current in the primary coil, $h_P$ ($h_S$) is the distance from the primary (secondary) coil to the sample, G is the total conductance of the sample, and M(x) is a geometrical factor [12]. Since the total impedance of the sample is given by [15] $Z = R + i\omega L_K$ the expression for the sample's total conductance reads:

$$G = \frac{1}{R + i\omega L_K} \qquad (2)$$

Here $L_k$ and R are the kinetic inductance and the resistance of the sample, respectively. From the above equations it follows that by measuring $V_{AC}(T)$ we can numerically reproduce the temperature dependencies of both $L_k$ and R.

From the two-fluid model, the relation between $L_k$ and $\lambda(T)$ for thin films (with thickness $d \ll \lambda$) is given by [1,2,15]:

$$L_K = \mu_0 \lambda \coth\left(\frac{d}{\lambda}\right) = \mu_0 \lambda \left(\frac{\lambda}{d}\right) \qquad (3)$$

This expression will be used hereafter to obtain λ(T) from the measured $L_K$(T) dependence. Instead of the tabulation based procedure used before [12], in the present study we have simultaneously determined G(T) from Eq.(1) and extracted both R(T) and $L_K$(T) using Eq.(2). Then from the temperature dependence of $L_K$ we recover the temperature dependence of λ. Typical results for extracted variation of $\lambda^2(T)/\lambda^2(0)$ for both SCCO and PCCO thin films are shown in Fig. 4.

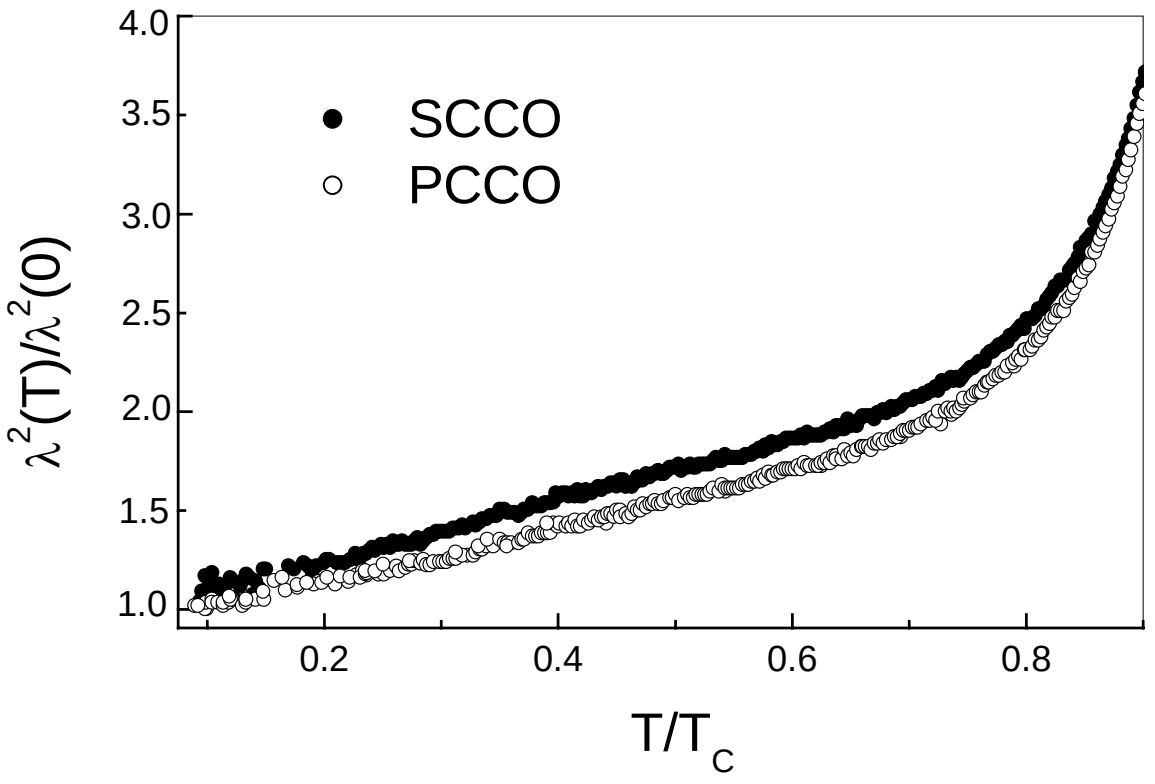

**Figure 4** – Extracted variation of $\lambda^2(T)/\lambda^2(0)$ as a function of the reduced temperature, obtained from Eqs.(1)-(3) for PCCO ($T_C$ =22.4 K) and SCCO ($T_C$ =20.2 K) thin films.

Turning to the discussion of the obtained results, recall [1] that for conventional BCS-type superconductors with *s*-wave pairing symmetry the superfluid fraction $x_S(T) = \lambda^2(0)/\lambda^2(T)$ saturates exponentially as T approaches zero. On the other hand, for a superconductor with a line of nodes, $x_S(T)$ will show a power-like behavior at low temperatures. In particular, the simple $d_{x^2-y^2}$ pairing state gives a linear dependence [16] $\Delta\lambda(T)/\lambda(0) \propto T$ for the low-temperature variation of in-plane penetration depth $\Delta\lambda(T)=\lambda(T)-\lambda(0)$. At the same time, in the presence of strong enough impurity scattering the linear T dependence changes to a quadratic $T^2$ dependence [17-20] $\Delta\lambda(T)/\lambda(0) \propto T^2$.

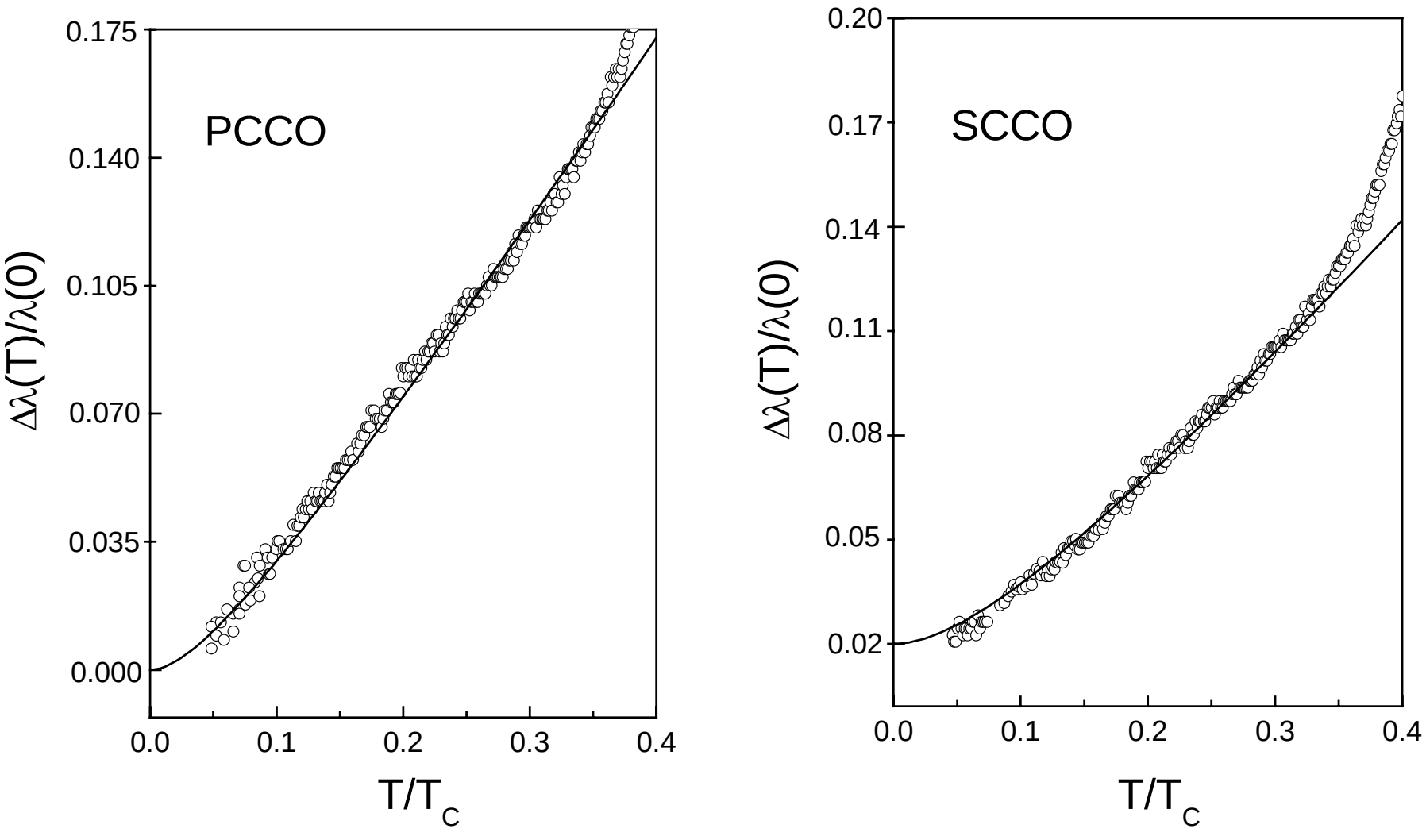

**Figure 5** – Low temperature fits (solid lines) of the extracted variation of the penetration depth $\Delta\lambda(T)/\lambda(0)$ in PCCO (left) and SCCO (right) thin films using the Goldenfeld-Hirschfeld interpolation formulae.

By trying many different temperature dependencies (including both exponential and power-like), we found that both our samples are best-fitted (see Fig. 5) by the so-called Goldenfeld-Hirschfeld interpolation formulae [20] $\Delta\lambda(T)/\lambda(0)=AT^2/(T+T_0)$ which describes a crossover between linear and quadratic dependencies above and below some temperature $T_0$. Here $A=\ln(2)k_B/\Delta_0$ with $\Delta_0$ being the amplitude of the zero-temperature value of the *d*-wave gap parameter, and the crossover temperature $T_0$ depends on the (unitary limit) scattering rate $\Gamma$ (which is proportional to the impurity concentration of the sample) as follows $T_0=\ln(2)k_B\Gamma^{1/2}\Delta_0{}^{1/2}$. The fitting parameters are given in Table 1. Noticeably, the crossover temperature $T_0$ is lower for high-quality PCCO films ($T_0/T_C$ =0.13). In turn, this observation correlates well with a lower value of impurity scattering rate (in dimensionless units, $k_B{}^3\Gamma/T_C$=0.017). Notice that the above estimates are in good agreement with the known results for high-quality PCCO thin films [8]. On the other hand, a less homogeneous SCCO film (with $T_C$ =20.2K) exhibits a much stronger impurity scattering with the rate $k_B{}^3\Gamma/T_C$=0.062 (starting to dominate below $T_0/T_C$ =0.26). And finally, it is worth mentioning that the reported in the literature [10] enhancement of diamagnetic screening in SCCO below 4K (related to antiferromagnetic ordering) is not seen in our thin films most probably due to insufficient structural homogeneity of our samples (see Figs. 1 and 2).

## 4. Conclusion

In summary, by using a highly-sensitive home-made mutual-inductance technique associated with a new numerical procedure, we extracted with high accuracy the temperature profiles of penetration depths $\lambda(T)$ in optimally-doped $Pr_{1.85}Ce_{0.15}CuO_4$ (PCCO) and $Sm_{1.85}Ce_{0.15}CuO_4$ (SCCO) thin films. Based on the obtained results, we can conclude that, first of all, our findings confirm a universal pairing mechanism in electron-doped magnetic superconductors with *d*-wave nodal symmetry and secondly, that deviations from the expected wave symmetry at the lowest temperatures are clear signals of structural inhomogeneity which can be tested via accurate measurement of the magnetic penetration depth. More precisely, both samples were found to follow the Goldenfeld-Hirschfeld interpolation formulae $\Delta\lambda(T)/\lambda(0)=AT^2/(T+T_0)$ with $T_0$ being the crossover temperature which demarcates pure and impure scattering contributions. The values of the extracted impurity scattering rate $\Gamma$ were found to correlate with the quality of our samples. As expected, small (large) values of $\Gamma$ correspond to high (low) values of the critical temperature $T_C$ in more (less) homogeneous PCCO (SCCO) thin films.

## Acknowledgements

We gratefully acknowledge financial support from Brazilian agency FAPESP. We also thank S. Anlage, C. J. Lobb and R. L. Greene from the *Center for Superconductivity Research* (University of Maryland at College Park) for useful comments and discussions.

**Table 1.** Fitting parameters for temperature dependencies of penetration depth variations $\Delta\lambda(T)/\lambda(0)$ extracted from PCCO and SCCO thin films (see Fig. 5) along with the estimates for the nodal gap parameter $\Delta_o$ and impurity scattering rate $\Gamma$ (in dimensionless units).

| sample | $T_C$ (K) | A $T_C$ | $T_o/T_C$ | $\Delta_o/k_BT_C$ | $\Gamma k_B^3/T_C$ |
|---|---|---|---|---|---|
| PCCO | 22.4 | 0.35 | 0.13 | 2.0 | 0.017 |
| SCCO | 20.2 | 0.33 | 0.26 | 2.1 | 0.062 |